\documentclass[aps,pre,reprint,final,10pt,twocolumn,oneside,nobalancelastpage,
flushbottom,groupedaddress,superscriptaddress,showpacs,showkeys,
amsfonts,amssymb,amsmath,longbibliography]{revtex4-1}
\usepackage[T2A]{fontenc}
\usepackage[utf8]{inputenc}
\usepackage[left=2cm,right=1.5cm,top=1.5cm,bottom=2cm]{geometry}
\usepackage{color,xcolor}
\usepackage{graphicx}%\graphicspath{{Figures/}}
\usepackage{epstopdf}
\usepackage{hyperref}
\usepackage{cleveref}
\usepackage{natbib}
\usepackage{bm}

\begin{document}



\title{Symmetry broken states in an ensemble of globally coupled pendulums}

\author{Maxim~I.~Bolotov}
\affiliation{Department of Control Theory, Nizhny Novgorod State University, Gagarin Av. 23, Nizhny Novgorod, 603950, Russia}

\author{Vyacheslav~O.~Munyaev}
\affiliation{Department of Control Theory, Nizhny Novgorod State University, Gagarin Av. 23, Nizhny Novgorod, 603950, Russia}

\author{Lev~A.~Smirnov}
\affiliation{Department of Control Theory, Nizhny Novgorod State University, Gagarin Av. 23, Nizhny Novgorod, 603950, Russia}
\affiliation{Institute of Applied Physics, Russian Academy of Sciences, Ul’yanova Str. 46, Nizhny Novgorod, 603950, Russia}

\author{Alexander~E.~Hramov}
\affiliation{Innopolis University, Universitetskaya Str.~1, Innopolis, 420500, Russia}

\begin{abstract}
We consider the rotational dynamics in an ensemble of globally coupled identical pendulums. This model is essentially a generalization of the standard Kuramoto model, which takes into account the inertia and the intrinsic nonlinearity of the community elements. There exists the wide variety of in-phase and out-of-phase regimes. Many of these states appear due to broken symmetry. In the case of small dissipation our theoretical analysis allows one to find the boundaries of the instability domain of in-phase rotational mode for ensembles with arbitrary number of pendulums, describe all arising out-of-phase rotation modes and study in detail their stability. For the system of three elements parameter sets corresponding to the unstable in-phase rotations we find a number of out-of-phase regimes and investigate their stability and bifurcations both analytically and numerically. As a result, we obtain a sufficiently detailed picture of the symmetry breaking and existence of various regular and chaotic states.
\end{abstract}
	\date{\today}
\date{\today}

\maketitle
%------------------------------------------------------------------------------%
\section{Introduction}\label{sec:Introduction}
%------------------------------------------------------------------------------%
The study of collective dynamics in networks of coupled oscillatory elements is one of the most attracting topics in modern nonlinear dynamics. It is important both for theoretical and practical points of view. A general phenomenon of collective behavior is synchronization~\cite{Pikovsky, Osipov, Afraimovich, Mosekilde, Anishchenko, Balanov}. Synchronization is usually understood as a process of achieving the collective rhythm of functioning by coupled objects of different nature. Synchronization of two or three elements is possible, as well as of ensembles consisting of hundreds and thousands of elements~\cite{Pikovsky, Osipov, Afraimovich, Mosekilde, Anishchenko, Balanov}. Even a weak attracting coupling
can adjust phases and frequencies of oscillators, and they can synchronize. Now three types of synchronization in its networks are known: full (or global) synchronization, partial (or cluster) synchronization and chimera states.

The system of coupled pendulums is one of the widely used models in multiple fields of science and technology. Despite the simplicity of this model, it adequately describes not only mechanical objects~\cite{Kecik}, but also various processes that occur in semiconductor structures~\cite{Barone}. This model is often considered as the basis for the theoretical studies of coupled Josephson junctions~\cite{Pikovsky, Barone}. Hence, it is very important to investigate the behavior of this system.

Cluster and {solitary} states in ensembles of coupled elements are of special interest for an investigation of the synchronization and symmetry breaking phenomena~\cite{PikovskyRosenblum2015, Kurths2016}. The first type states consist of two or more groups in which an individual oscillators behave identically. These states have been well known for many years, but have still attracted great attention of investigators across different fields of science and engineering~\cite{PikovskyRosenblum2015, Kurths2016}. {Cluster states arise both in ensembles with a finite number of elements~\cite{Kaneko1990, Okuda1993, Nakagawa1994,Schmidt2014}, and in distributed oscillatory media~\cite{Vanag2000,Mikhailov2006,Lin2004}. In the article~\cite{Kemeth2019} two-cluster regimes in small ensembles of Stuart-Landau oscillators are considered. Solitary state can be attributed to a special type of cluster state. Solitary state is formed when a single oscillator is separated from a synchronous cluster in an ensemble~\cite{Maistrenko2014}. Examples of solitary states were found in various types of networks~\cite{Mikhaylenko2019,Majhi2019}, systems of non-locally coupled elements~\cite{Jaros2015}, systems of Stuart-Landau oscillators~\cite{Schmidt2014}.}

In this paper we consider features found in rotational dynamics of ensemble of globally coupled pendulums. Note that one can interpret our basic model as a generalization of the standard Kuramoto model, which takes into account the inertia and the intrinsic nonlinearity of the elements of the discussed population. The influence of the first type effects have been intensevely studied in recent years (e.g., see~\cite{Ji,Ha,Belykh2016}). Actually, the modifications of the second type have been also considered in the literature~\cite{Komin,Daido,Lafuerza}. In this work, we assume, that both effects are present and play an important role. Hence, the Kuramoto model describing the evolution of a group of phase oscillators with global coupling transforms to the system, consisting of a number of pendulums with meanfield interaction. We are interested in in-phase rotations and nontrivial out-of-phase ones. In the present article, we restrict ourselves to considering ensembles with a small number $N$ of elements and focus on all the possible limit motions of the system.

This paper is organized as follows. In Section~\ref{sec:Model}, we describe the model, state the problem. In Section~\ref{sec:InPhaseMode} we report on the numerically observed effect: in-phase periodic motion instability. Using the assumption of small dissipation we developed an asymptotic theory, which explains instability of the in-phase periodic motion (essentially presented limit cycle on the cylinder) of the pendulums. Here we also find analytical expression for the boundaries of the in-phase limit rotation mode instability interval regarding the coupling strength. During the nonlinear stage of this instability a periodic out-of-phase rotation emerges, in particular, a solitary state for which the phases of some pendulums coincide, while the phases of the other pendulums differ from the rest. Section~\ref{sec:ParametricInstability} contains a detailed description of the general analytical approach for stability analysis of out-of-phase periodic limit cycles existing in the ensemble of globally coupled pendulums.
In Section~\ref{sec:LimitCycles} theoretical and numerical results are presented.
In particular,in Section~\ref{sec:Stability_Out_Of_Phase} the main results for stability analysis of out-of-phase rotation modes within the framework of the considered model are presented.
In Section~\ref{sec:Ring} bifurcations that lead to the appearance and disappearance of the out-of-phase limit rotation modes are analyzed. Bistability of the in-phase and out-of-phase limit periodic modes is established for the system under study.
In Section~\ref{sec:Chaos}, scenario of chaotic rotational dynamics emergence is described.
In Section~\ref{sec:N_4}, examples of out-of-phase regimes, existing in larger ensembles, are considered. A summary of the main results can be found in Conclusion. Appendix~\ref{sec:Stability_Analysis} contains the details about the linear stability analysis for out-of-phase rotation modes.
In Appendix~\ref{sec:Numerical_Setup} we present a brief description of the numerical methods used for calculating any possible periodic modes and their linear stability within the framework of the considered model.

%------------------------------------------------------------------------------%
\section{Nature of symmetry breaking in a system of coupled pendulums
}\label{sec:MainSection}
%------------------------------------------------------------------------------%
\subsection{Model under study}\label{sec:Model}
%------------------------------------------------------------------------------%
We study the rotational dynamics of an ensemble of $N$ globally coupled identical pendulum-like oscillators (index $n$). The setup we employ can be considered as a generalization of the standard Kuramoto model, which takes into account the inertia and the intrinsic nonlinearity of the structural elements. The nonlinearity in the form of sinus was chosen because it is rather typical for many real systems indicated above. In this case, the evolution of the generalized coordinate $\varphi_n(t)$ of the $n$-th unit belonging to such a community is given by
\begin{equation}
\ddot{\varphi}_n+\lambda \dot{\varphi}_n+\sin\varphi_{n}=\gamma+\dfrac{K}{N} \sum_{\tilde{n}=1}^{N}\sin\left(\varphi_{\tilde{n}}-\varphi_{n}\right),
\label{eq:EqPhi1}
\end{equation}
where $\lambda$ is the damping coefficient responsible for all the dissipative processes in the system, $\gamma$ is a constant external force identical for all pendulums, $K$ characterizes the strength of global coupling between the oscillators in the discussed group.

{Since we are interested in the rotational dynamics of system~\eqref{eq:EqPhi1}, we begin with considering the simplest in-phase rotational mode and its stability.} When at any given time $t$ coordinates $\varphi_{n}(t)$ and velocities $\dot{\varphi}_{n}(t)$ $(n=1,\dots,N)$ coincide, the system demonstrates in-phase dynamics, i.e. at any time $t$: $\varphi_{1}\left(t\right)=\dots=\varphi_{N}\left(t\right)=\phi\left(t\right)$. {We denote such a regime as $(N:0)$.} In this regime all pendulums move synchronously and their dynamics is described by the following single equation:
\begin{equation}
\ddot{\phi}+\lambda \dot{\phi}+\sin\phi=\gamma,
\label{eq_pendula}
\end{equation}
which essentially represents a nonlinear damped pendulum equation with a constant external force. The dynamical behaviour, all possible equilibrium states and limit motions of a particle that is governed by Eq.~(\ref{eq_pendula}) have been well studied and thoroughly analysed in many publications (e.g., see \cite{Andronov, Tricomi1933, Belykh1977, Peng} for details). Based on the previous results, here we briefly outline the main specific features of the model~(\ref{eq_pendula}), that make the formulation of the problem investigated in the presented paper clear and accessible to the broad readership.
The parameter plane $\lambda$, $\gamma$ is divided into three domains \cite{Tricomi1933, Belykh1977, Peng, Strogatz}, which correspond to different structurally stable cylindrical phase spaces of the system~(\ref{eq_pendula}). For $\lambda$, $\gamma$ from the first domain, only two steady states, namely, a saddle and a focus (node), exist in the phase space of the corresponding pendulum. When $\lambda$, $\gamma$ are from the second domain in addition to these steady states, there is also a stable $2\pi$-periodic in $\phi$ limit cycle. Noteworthy, in this case, the attraction region of a focus (node) is small enough and is delimited by separatrices of a saddle. For $\lambda$, $\gamma$ from the third domain, equilibrium states disappear from the system, and there remains only an attractive rotation mode. The third domain is separated from the others by the straight line $\gamma\!=\!1$ in the parameter plane $\lambda$, $\gamma$. In turn, the first and second domains are separated by the so-called Tricomi bifurcations curve~\cite{Tricomi1933, Belykh1977, Peng, Strogatz}.

In the case of small dissipation (i.e. when $\lambda\!\ll\!\gamma$), it is possible to analytically describe a rotational limit motion of one pendulum, using an asymptotic approach based on, e.g. the Lindstedt-Poincar\'e method \cite{Nayfeh}. It has been shown in \cite{Bolotov2019} that an approximation for limit rotation solution of the system~\eqref{eq_pendula} has the following form:
\vspace{-.1in}
\begin{equation}
\phi(\tau)\!=\!\tau\!+\!\dfrac{\lambda^2}{\gamma^2}\sin\tau\!+\!o\!\left(\lambda^4\right),\hspace{2mm}
\tau \!=\! \left(\dfrac{\gamma}{\lambda} \!-\! \dfrac{\lambda^3}{2\gamma^3} \!+\! o\!\left(\lambda^7\right)\right) t.
\label{eq_rot_cycle_phase}
\end{equation}
We note, the system~\eqref{eq_pendula} has a continuum of oscillatory (not rotatory) type solutions at $\lambda = 0$ and $\gamma \in [0,1) $. If $\lambda$ is not equal to 0, the stable steady state (center) is transformed to the stable focus and all these solutions do not exist and only one periodic rotatory solution can appear.

%------------------------------------------------------------------------------%
\subsection{Stability of in-phase rotation mode}\label{sec:InPhaseMode}
%------------------------------------------------------------------------------%
The in-phase rotation mode is stable in the wide range of the parameters $\lambda$, $\gamma$ and $K$, according to our numerical simulations directly in the framework of the basic model~(\ref{eq:EqPhi1}). However, for certain values of $\lambda$, $\gamma$ and $K$ the system~(\ref{eq:EqPhi1}) demonstrates non-trivial behaviour: the instability of in-phase rotation limit motion is developed and, as a result, one of a variety of out-of-phase states appears, that leads to partially broken symmetry in the considered ensemble of globally coupled pendulums. It is worth mentioning that similar in nature effect takes place for the system of only two elements~\cite{Smirnov2016}. This effect is also observed in the case of a small chain of nearest-neighbour interacting identical pendulums \cite{Bolotov2019}.

Before proceeding to a detail description of the variety of out-of-phase rotation modes observed in direct numerical simulations within the model~(\ref{eq:EqPhi1}), we propose an analytical approach, which permits one to explain an instability of periodic limit cycles existing in the system of globally coupled pendulums.
As a starting point, we consider a possibility of developing of such an instability process for the in-phase state discussed above. This allows us to demonstrate the basic idea of our theoretical analysis while avoiding cumbersome calculations.

Let us find the stability conditions for the in-phase rotational mode. First we linearize the system~(\ref{eq:EqPhi1}) around $\phi(t)$, then ${\varphi}_n(t) = \phi(t) + \delta {\varphi}_n(t)$ and $\dot{\varphi}_n(t) = \dot{\phi}(t) + \delta \dot{\varphi}_n(t)$, where perturbations $\delta {\varphi}_n(t)$, $\delta \dot{\varphi}_n(t)$ are small in magnitude. Next we get the corresponding equations for variations $\delta {\varphi}_n(t)$:
\begin{equation}
\begin{gathered}
\delta\ddot{\varphi}_n + \lambda \delta\dot{\varphi}_n + \cos \phi(t) {\delta\varphi}_n = \dfrac{K}{N} \sum_{\tilde{n}=1}^{N}({\delta\varphi}_{\tilde{n}} - {\delta\varphi}_n).
\end{gathered}
\label{eq_var}
\end{equation}
To continue with Eqs.~\eqref{eq_var}, we introduce new variables
\begin{equation}
\begin{gathered}
\eta = \frac{1}{N}\sum_{\tilde{n}=1}^{N}\delta\varphi_{\tilde{n}},\\
\xi_n = \delta\varphi_{n+1} - \delta\varphi_{n},\hspace{2mm} n = 1, \dots, N-1,
\end{gathered}
\label{eq_subst_dir}
\end{equation}
for which Eqs.~\eqref{eq_var}, corresponding to the in-phase $T$-periodic rotational cycle of the system~\eqref{eq:EqPhi1}, take the following form
\begin{equation}
\begin{gathered}
\ddot{\eta} + \lambda \dot{\eta} + \cos \phi(t) \eta = 0,
\label{eq_phaseDiffN_1}
\end{gathered}
\end{equation}
\begin{equation}
\begin{gathered}
\ddot{\xi}_{n} + \lambda \dot{\xi}_{n} + (K + \cos \phi(t)) \xi_{n} = 0,\\
n=1, \dots, N-1.
\end{gathered}
\label{eq_phaseDiffN_2}
\end{equation}
The Floquet multipliers of Eq.~\eqref{eq_phaseDiffN_1} are equal to $1$ and $e^{-\lambda T}$. Indeed, differentiation of Eq.~\eqref{eq_pendula} with respect to $t$ shows that periodic function $\dot{\phi}\left(t\right)$ is the solution to Eq.~\eqref{eq_phaseDiffN_1}, so the corresponding multiplier is equal to 1. The second multiplier can be found using Liouville's formula. Therefore $\eta$ mode is always stable, so instability in the system can arise only due to the excitation of modes $\xi_n$ $(n = 1, \dots, N - 1)$. Eq.~\eqref{eq_phaseDiffN_2} belongs to the Mathieu-type equation. Hence, the parametric instability effects can be observed for some values of the parameter $K$ depending on $\lambda$ and $\gamma$ \cite{McLachlan}. To find the boundaries of the instability domain of the in-phase rotation mode, we determine the coupling parameter $K$ values for which the Eq.~(\ref{eq_phaseDiffN_2}) admits a solution with period $2T$ or, equivalently, with frequency $\omega / 2$. Passing to a dimensionless time $t = \omega \tau$, we reduce the system~(\ref{eq_phaseDiffN_2}) to the following form
\begin{equation}
\begin{gathered}
\ddot{\xi}_{n} + {\lambda}{\omega} \dot{\xi}_{n} + {\omega^2}\left[{K} + \cos(\phi(\tau))\right]{\xi}_{n} = 0,\\
n=1, \dots, N-1.
\end{gathered}
\vspace{-.1in}
\label{eq_1+1+1_SmallPar}
\end{equation}
Using the perturbation theory, taking result \eqref{eq_rot_cycle_phase} and searching for a solution to Eq.~\eqref{eq_1+1+1_SmallPar} with $\omega / 2$ frequency, we get boundaries $K_{1,2}$ for the instability
domain
\begin{equation}
\begin{gathered}
K_{1,2} = \dfrac{1}{4} \left[\dfrac{\gamma^2}{\lambda^2} \mp 2 \sqrt{1 - \gamma^2} + \dfrac{1}{2} \dfrac{\lambda^2}{\gamma^2}\right] + O(\lambda^4).
\label{eq_alpha0_K12_1}
\vspace{-.1in}
\end{gathered}
\end{equation}
When the critical values of $K_{1,2}$ are reached, the instability of the $\xi_{n}$, $n=1, \dots, N-1$ modes simultaneously develops, which is a common property at destabilization of the synchronous states in globally coupled systems of $N$ identical units. Thus, in the system of globally coupled pendulums \eqref{eq:EqPhi1}  only one instability region ($K_1 < K < K_2$) of the in-phase rotational motion $\phi(t)$ exists, which boundaries $K_1$ and $K_2$ are determined by the expression \eqref{eq_alpha0_K12_1}.

\begin{figure}[t]
	\centering
	\includegraphics[width=1.0\columnwidth]{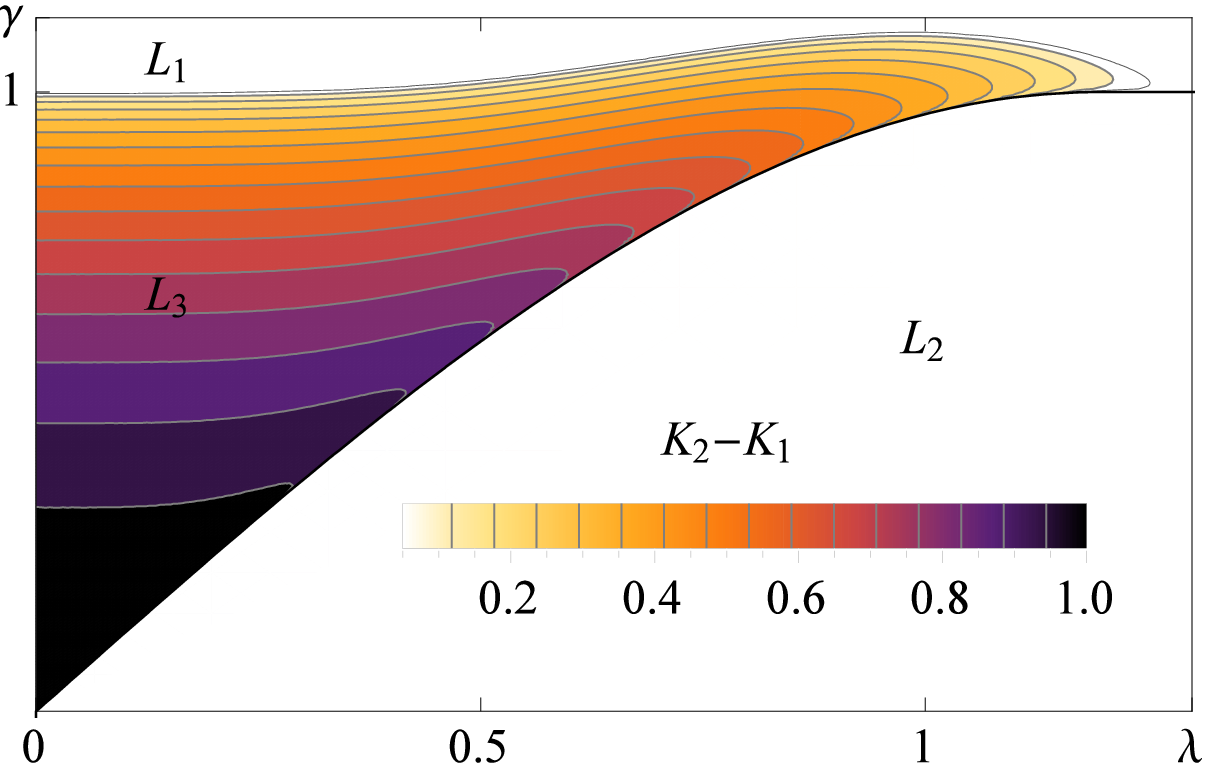}	
	\caption{\label{fig:InstabAreaWidth} The range of values of coupling strength $K$, when in-phase mode instability is observed, depending on the values of parameters $\gamma$ and $\lambda$ as a function $[K_2 - K_1](\lambda, \gamma)$. The black curve is the Tricomi bifurcation curve, i.e. boundary of the in-phase rotational mode existence region. For region $L_1$ in-phase rotation is stable with any $K$. In region $L_2$ in-phase rotation does not exist. In-phase rotation in region $L_3$ is unstable for $K_1 < K < K_2$.}
\end{figure}

Fig.~\ref{fig:InstabAreaWidth} shows the width {of the range of values of coupling strength $K$, when in-phase mode instability is observed, depending on the values of parameters $\gamma$ and $\lambda$ as a function $[K_2 - K_1](\lambda, \gamma)$. The parameter plane $(\lambda, \gamma)$ can be divided into 3 areas: $L_1$, $L_2$ and $L_3$. In the area $L_1$, the in-phase mode is stable for any $K$. In the area $L_2$, bounded by the Tricomi curve and line $\gamma = 1$, this regime does not exist. The Tricomi curve corresponds to the values of the parameters $(\lambda, \gamma)$, for which the rotational motion in the system~\eqref{eq_pendula} disappears as a result of the saddle-node homoclinic bifurcation. In the color region $L_3$ there is a range of values of $K$ for which the in-phase rotation is unstable ($K_1 < K <K_2$). The boundaries $K_1$ and $K_2$ of this range are determined by expressions~\eqref{eq_alpha0_K12_1}. Black color means, that the length of the instability interval of $(N:0)$ regime is close to one. Orange color means, that length of the instability interval of $(N:0)$ regime is close to zero. By the fixed $\lambda$ with increase of $\gamma$ the interval $K_2-K_1$ becomes smaller. The region of instability arises at values $\gamma$ close to unity and becomes wider as $\gamma$ decreases. When the parameters $(\gamma, \lambda)$ reach the Tricomi curve, the in-phase rotational mode disappears.}

%------------------------------------------------------------------------------%
\subsection{General approach for stability of analysis of symmetry broken states }\label{sec:ParametricInstability}
%------------------------------------------------------------------------------%
Let us proceed to a detailed description of the general analytical approach for stability analysis of out-of-phase periodic limit cycles existing in the model (\ref{eq:EqPhi1}) of $N$ globally coupled pendulums. Just for definiteness assume that pendulums form $M$ synchronous clusters with $N_m$ pendulums in $m$-th cluster ($N_1+\ldots+N_M = N$). Here and below such regimes will be denoted as $\left(N_1:N_2:\ldots:N_M\right)$. Note that a cluster also may consist of only one element. Assuming what every pendulum in the $m$-th cluster is moving with the same phase $\phi_m\left(t\right)$ and velocity $\dot{\phi}_m\left(t\right)$ the system (\ref{eq:EqPhi1}) of $N$ second order equations reduces to the set of $M$ second order differential equations which defines dynamics of the unknown phases $\phi_m\left(t\right)$:
\vspace{-.1in}
\begin{equation}
\ddot{\phi}_m+\lambda \dot{\phi}_m+\sin\phi_{m}=\gamma+\dfrac{K}{N} \sum_{\tilde{m}=1}^{M}N_{\tilde{m}}\sin\left(\phi_{\tilde{m}}-\phi_{m}\right).
\label{eq:EqPhi2}
\vspace{-.1in}
\end{equation}
Applying the same approach described in Section~\ref{sec:Model} for the in-phase solution of (\ref{eq:EqPhi1}) to the current problem of finding $T$-periodic limit cycles of the Eq.~\eqref{eq:EqPhi2} the phases $\phi_m\left(t\right)$ can be determined.

After phases $\phi_m\left(t\right)$ are found it is possible to proceed with stability analysis of the regime $\left(N_1:N_2:\ldots:N_M\right)$. For that one imposes a linear perturbation $\varphi_{m k}\left(t\right) = \phi_{m}\left(t\right) + \delta\varphi_{m k}\left(t\right)$ and $\dot{\varphi}_{m k}\left(t\right) = \dot{\phi}_{m}\left(t\right) + \delta\dot{\varphi}_{m k}\left(t\right)$, where $\varphi_{m k}\left(t\right)$, $\dot{\varphi}_{m k}\left(t\right)$ are the perturbed phase and velocity of $k$-th pendulum from the $m$-th cluster ($k=1,\ldots,N_m$). The equations for variations $\delta\varphi_{m k}\left(t\right)$ thus read
\begin{equation}
\begin{gathered}
\delta\ddot{\varphi}_{m k}+\lambda \delta\dot{\varphi}_{m k}+\cos\phi_m\delta\varphi_{m k}\\
=\dfrac{K}{N} \sum_{\tilde{m}=1}^{M}\sum_{\tilde{k}=1}^{N_{\tilde{m}}}\cos\left(\phi_{\tilde{m}}-\phi_m\right)\left(\delta\varphi_{\tilde{m} \tilde{k}}-\delta\varphi_{m k}\right).
\label{eq:EqDeltaPhi}
\vspace{-.1in}
\end{gathered}
\end{equation}
The next key step is the linear change of variables which breaks system \eqref{eq:EqDeltaPhi} into a set of independent differential equations systems of smaller dimensions:
\vspace{-.1in}
\begin{equation}
\begin{gathered}
\eta_m = \frac{1}{N_m} \sum_{k=1}^{N_m} \delta\varphi_{m k}, \hspace{2mm} m = 1, \ldots ,M, \\
\xi_{m k} = \delta\varphi_{m, k + 1} - \delta\varphi_{m k}, \hspace{2mm} k = 1, \ldots ,N_m-1.
\label{eq:subst_dir}
\vspace{-.1in}
\end{gathered}
\end{equation}
Excitation of mode $\eta_{m}$ in the case $\xi_{\tilde{m} \tilde{k}} = 0$ can be interpreted as perturbation of $m$-th cluster like whole unit, because in this case all pendulums forming cluster $m$ are identically perturbed. Stability associated with such kind of perturbations are often called internal stability \cite{LyapunovExponents}. On the contrary excitation of modes $\xi_{m k}$ for fixed $m$ in the case $\eta_{\tilde{m}} = 0$ means arbitrary perturbation of pendulums only inside the $m$-th cluster and associated stability is called external stability \cite{LyapunovExponents}.

For new variables Eqs. \eqref{eq:EqDeltaPhi}, corresponding to the $T$-periodic rotational cycle of the system (1), take the following form
\begin{equation}
\begin{gathered}
\ddot{\eta}_m+\lambda \dot{\eta}_m+\cos\phi_m\eta_m\\
=K \sum_{\tilde{m}=1}^{M}\dfrac{N_{\tilde{m}}}{N}\cos\left(\phi_{\tilde{m}}-\phi_m\right)\left(\eta_{\tilde{m}}-\eta_m\right),
\vspace{-.1in}
\label{eq:EqEta}
\end{gathered}
\end{equation}
\begin{equation}
\ddot{\xi}_{m k}+\lambda\dot{\xi}_{m k}+\left(\cos\phi_m + K\sum_{\tilde{m}=1}^{M}\frac{N_{\tilde{m}}}{N}\cos\left(\phi_{\tilde{m}}-\phi_m\right)\right)\xi_{m k}\\
=0.
\label{eq:EqXi}
\end{equation}
So actually it is necessary to analyze the system of $M$ second order equations \eqref{eq:EqEta} and {$\tilde{M}$} independent equations of second order \eqref{eq:EqXi} (because of their similarity for different $k$, so index $k$ will be omitted below), {where $\tilde{M} = \left|\left\{m : N_m > 1\right\}\right|$ is the number of clusters in which the number of elements is greater than one}.

Floquet's Theorem implies that to each characteristic exponent $\Lambda$ of the linear differential equations system with periodic coefficients, such as Eqs. \eqref{eq:EqEta} is, there corresponds a solution of the form $\eta_m(t)=e^{\Lambda^{in} t} \tilde{\eta}_m(t)$ ($m=1,\ldots,M$), where $\tilde{\eta}_m(t)$ is periodic in $t$: $\tilde{\eta}_m\left(t+T\right)=\tilde{\eta}_m\left(t\right)$. New variables $\tilde{\eta}_m(t)$ satisfy the equations:
\begin{equation}
\begin{gathered}
\ddot{\tilde{\eta}}_m+\left(2 \Lambda^{in}+\lambda\right) \dot{\tilde{\eta}}_m+\left(\left(\Lambda^{in}\right)^2+\lambda\Lambda^{in}+\cos\phi_m\right)\tilde{\eta}_m\\
=K\sum_{\tilde{m}=1}^{M}\dfrac{N_{\tilde{m}}}{N}\cos\left(\phi_{\tilde{m}}-\phi_m\right)\left(\tilde{\eta}_{\tilde{m}}-\tilde{\eta}_m\right).
\vspace{-.1in}
\label{eq:EqTildeEta}
\end{gathered}
\end{equation}

To determine the characteristic exponent $\Lambda$, we use the Lindstedt-Poincar\'e method. First, since our theory is being developed near the conservative limit, i.e. for small dissipation $\lambda$, the  characteristic exponent $\Lambda$ is represented in the form of the asymptotic expansion
\begin{equation}
\Lambda^{in} = \sum_{j=1}^{+\infty} \lambda^j \Lambda^{(j)}.
\vspace{-.08in}
\label{eq:EqLambdaExpansion}
\end{equation}
Coefficients $\Lambda^{(j)}$ should be chosen so as to avoid secular terms in \eqref{eq:EqTildeEta}, i.e. so that $\tilde{\eta}_1,\ldots,\tilde{\eta}_M$ are periodic.

Equivalently, using Floquet's Theorem we are searching for the solution of \eqref{eq:EqXi} in the form $\xi_m(t)=e^{\Lambda^{ex} t}\tilde{\xi}_m(t)$, where $\tilde{\xi}_m(t)$ is periodic, i.e. $\tilde{\xi}_m\left(t+T\right)=\tilde{\xi}_m\left(t\right)$:
\vspace{-.1in}
\begin{equation}
\begin{gathered}
\ddot{\tilde{\xi}}_m+\left(2\Lambda^{ex}+\lambda\right)\dot{\tilde{\xi}}_m+\left(\left(\Lambda^{ex}\right)^2+\lambda\Lambda^{ex}+\cos\phi_m\right)\tilde{\xi}_m\\
=-K\sum_{\tilde{m}=1}^{M}\frac{N_{\tilde{m}}}{N}\cos\left(\phi_{\tilde{m}}-\phi_m\right)\tilde{\xi}_m.
\vspace{-.1in}
\label{eq:EqTildeXi}
\end{gathered}
\end{equation}
Using asymptotic expansion \eqref{eq:EqLambdaExpansion} and cancelling the secular terms in expression for $\tilde{\xi}_m$ unknown coefficients $\Lambda^{(j)}$ can be found.
\par
{The signs $\Lambda^{in}$ and $\Lambda^{ex}$ determine the stability of the modes $\eta_m$ and $\xi_m$ respectively. Thus obtaining the expression for $\phi_m(t)$ as an asymptotic expansion, similar to~\eqref{eq_rot_cycle_phase} in the case of in-phase rotations, we can get the decomposition in the form of a series for the Lyapunov exponents $\Lambda^{in}$ and $\Lambda^{ex}$ for the systems~\eqref{eq:EqTildeEta} and \eqref{eq:EqTildeXi}. In the end we have $2M$ exponents $\Lambda^{in}$ determining internal stability and $2(N-M)$ exponents $\Lambda^{ex}$ determining external stability. An example of symmetry broken states stability analysis is presented in~Sec.~\ref{sec:Stability_Out_Of_Phase}.}

%------------------------------------------------------------------------------%
\section{Theoretical and numerical results for the system of three elements}\label{sec:LimitCycles}
%------------------------------------------------------------------------------%
\subsection{Regimes (2:1) and (1:1:1) stability analysis}\label{sec:Stability_Out_Of_Phase}
%------------------------------------------------------------------------------%
In this section, we consider the application of the approach developed in the Sec.~\ref{sec:ParametricInstability} for analyzing the stability of cluster rotational motions resulting from the development of in-phase mode $\phi(t)$ instability for $N=3$. In this case, there are two types of cluster modes: (2:1) solitary state and (1:1:1) regime. Detailed stability analysis of these regimes is presented in Appendix~\ref{sec:Stability_Analysis}.

It is shown in \ref{sec:Regime_2_1} that for the solitary state (2:1) solutions $\phi_m^{-}$ arising on the right side $K_2$ of the instability interval of parameter $K$ defined by Eq.~\eqref{eq_alpha0_K12_1} is always internally unstable for all values of $K$ (see Eq.~\eqref{eq:2_1_Lambda_in_m}). Conversely, solution $\phi_m^{+}$ arising on the left side $K_1$ of the instability interval is always internally stable, but changes its external stability with $K=K_c^{(2:1)}$, and it is stable for $K > K_c^{(2:1)}$, where
\begin{equation}
\begin{gathered}
K_c^{(2:1)} = \dfrac{\gamma^2}{4 \lambda^2} + \dfrac{\sqrt{1-\gamma^2}}{3} + O(\lambda).
\label{eq_stability_2_1}
\end{gathered}
\end{equation}

Similarly for (1:1:1) regime from \eqref{eq:1_1_1_Lambda_ex_m} the instability of the solution $\phi_m^{-}$ for any value of $K$ follows. Solution $\phi_m^{+}$ changes its stability with $K=K_c^{(1:1:1)}$, and it is stable for $K < K_c^{(1:1:1)}$, where
\begin{equation}
\begin{gathered}
K_c^{(1:1:1)} = \dfrac{\gamma^2}{4 \lambda^2} + \dfrac{\sqrt{1-\gamma^2}}{3} + O(\lambda).
\label{eq_stability_1_1_1}
\end{gathered}
\end{equation}
{Thus, for certain values of the coupling strength $K$ the synchronous in-phase rotational mode in the ensemble of three coupled pendulums becomes parametrically unstable and, depending on the initial conditions, regimes (2:1) or (1:1:1) can be realized. The values of coupled strength parameters $K_c^{(2:1)}$ and $K_c^{(1:1:1)}$ at which the change of stability of cluster modes occurs were obtained.}
From numerical simulations it is known that $K_c^{(2:1)} \le K_c^{(1:1:1)}$ (see Section~\ref{sec:Ring}).

%------------------------------------------------------------------------------%
\subsection{Regular dynamics of in-phase and out-of-phase rotational modes}\label{sec:Ring}
%------------------------------------------------------------------------------%
In this and following section, we present the results of the numerical simulations (for details see Appendix~\ref{sec:Numerical_Setup}) which are performed directly within the framework of the discussed model~(\ref{eq:EqPhi1}) for a wide range of the parameters $\lambda$, $\gamma$ and $K$ in the case $N=3$. First of all, we consider in detail the development of the self-induced parametric instability of the in-phase synchronous regime and focus our attention on the nonlinear stage of this process and the resulting movements that can be set over long time. Our numerical calculations employed a commonly used fifth-order Runge-Kutta scheme (with fixed time step $dt=0.001$) to integrate the system~(\ref{eq:EqPhi1}).

\begin{figure}[h!]
	\centering
	\includegraphics[width=\columnwidth]{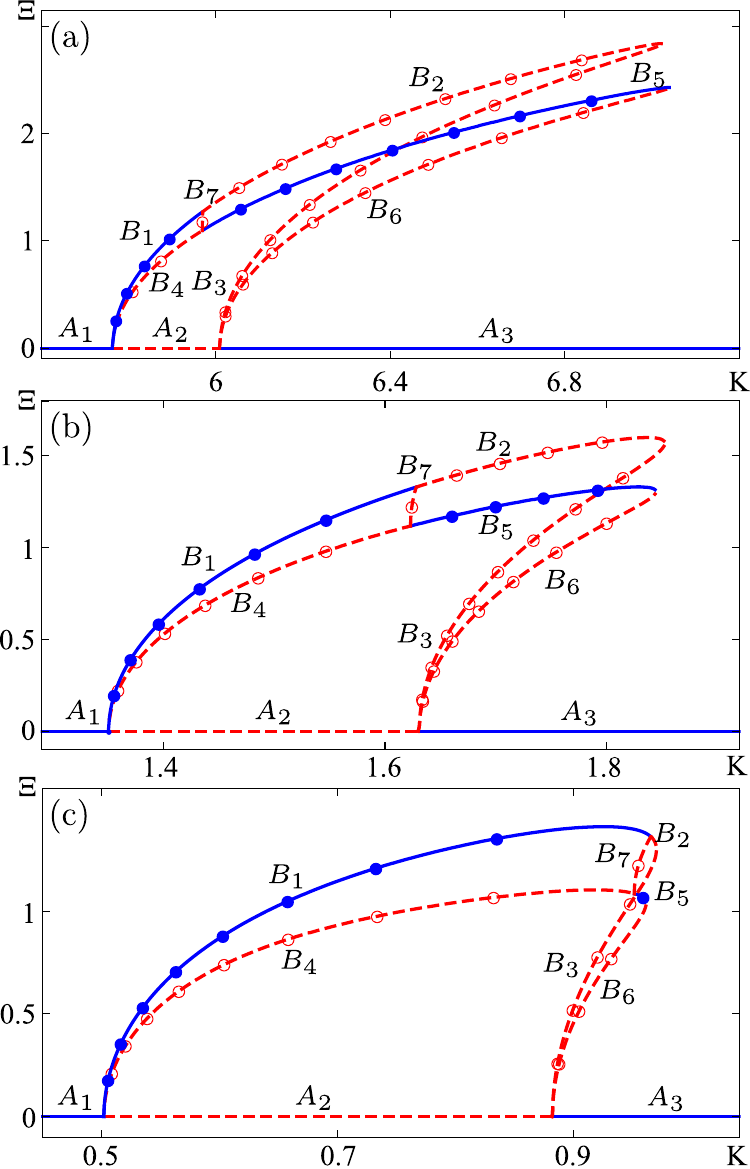}
	\caption{\label{fig4}~(Color online) Bifurcation diagram of synchronous rotational regimes of the system~\eqref{eq:EqPhi1} at $N=3$. Here and below: blue shared markers -- stable regimes, red unshared markers -- unstable regimes. Lines without markers -- $2\pi$-periodic regimes. Round markers -- $4\pi$-periodic regimes. Parameters: $\gamma = 0.97$. (a) $\lambda = 0.2$. (b) $\lambda = 0.4$. (c) $\lambda = 0.6$.}
\end{figure}

Let us consider the case $\gamma = 0.97$.
The bifurcation diagrams of evolution of periodical motions of the system~\eqref{eq:EqPhi1} are shown in Fig.~\ref{fig4}.
The diagram shows the dependence of synchronism characteristics $\Xi$ from magnitude of the coupling strength $K$.
First let us describe the diagram for $\gamma = 0.97$, $\lambda = 0.2$ (Fig.~\ref{fig4}a).
The horizontal segments $A_1$, $A_3$ correspond to the stable synchronous in-phase regime ($\Xi=0$).
There is a region $A_2$ of the parameter $K$ values, when this regime becomes unstable.
As shown above, in the course of the asymptotic consideration (the expression~\eqref{eq_alpha0_K12_1}), it is for the values of the coupling parameter $K$ that the parametric instability of the in-phase periodic motion develops from these intervals.
\par
\begin{figure}[h!]
	\centering
	\includegraphics[width=\columnwidth]{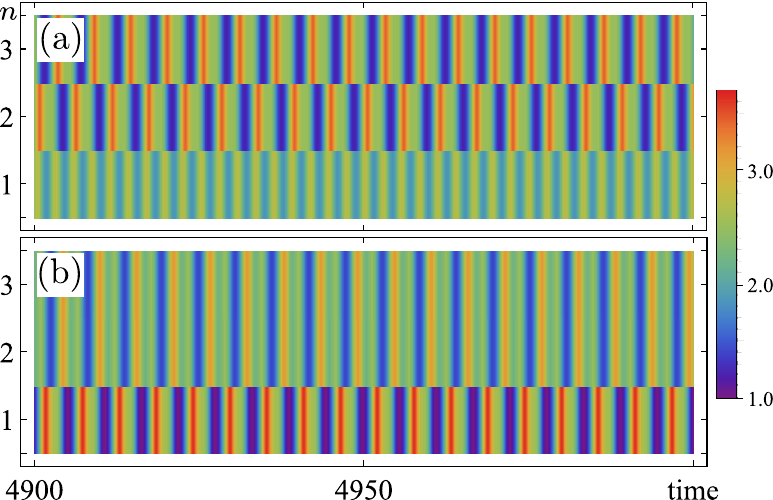}
	\caption{(Color online) Time dynamics of instantaneous frequencies $\dot{\varphi}_{n}$ ($n=1,2,3$) of the three pendulums in the system~\eqref{eq:EqPhi1} for $N=3$. (a) Regular (1:1:1) regime at $K = 1.6$. (b) Regular chimeric (2:1) regime at $K = 1.7$. Parameters: $\gamma = 0.97$, $\lambda = 0.4$.}
	\label{fig_regimes}
\end{figure}

Let us consider processes occurring in an ensemble when $K$ takes values from the $A_2$, and values outside it. 
As the parameter $K$ increases, the in-phase periodic motion undergoes period doubling bifurcation ($K \approx 5.764$), while from the stable $2 \pi$-periodic in $\varphi = (\varphi_1, \varphi_2, \varphi_3)^{T}$ rotation the stable (1:1:1) $4\pi$-periodic regime (branch $B_1$) and unstable $4 \pi$-periodic motion (2:1) (branch $B_4$) are generated, and $2\pi$-periodic regime loses stability.
In addition to the two above described $4\pi$-periodic motions, there are also two unstable out-of-phase $4\pi$-periodic (1:1:1) and (2:1) rotations in $\varphi$ (branches $B_3$ and $B_6$, respectively), which are generated from an unstable $2\pi$-periodic motion as a result of a subcritical period doubling bifurcation ($K \approx 6.008$) with increasing $K$. Corresponding values of the parameter $K$ for which doubling bifurcation occurs calculated analytically using Eq.~\eqref{eq_alpha0_K12_1} are $K \approx 5.764$ and $K \approx 6.007$.
The (1:1:1) regime (Fig.~\ref{fig_regimes}a) corresponds to the branches $B_1$, $B_2$, $B_3$, and the (2:1) solitary regime (Fig.~\ref{fig_regimes}b) corresponds to the branches $B_4$, $B_5$, $B_6$.
As the parameter $K$ increases the stability change of the (1:1:1) and (2:1) regimes is to see.
The first one loses stability at $K_c^{(1:1:1)} \approx 5.9699$, and the second one becomes stable at $K_c^{(2:1)} \approx 5.9687$ (according to theoretical predictions, see Eqs.~\eqref{eq_stability_2_1} and \eqref{eq_stability_1_1_1}, $K_c^{(1:1:1)} \approx K_c^{(2:1)} \approx 5.962$).
So there is a bistability region of (1:1:1) and (2:1) regimes.
The stability change occurs through the pitchfork bifurcation, and herewith the unstable (1:1:1) regime appears (brunch $B_7$).
Further with the increasing $K$ the periodic motions, corresponding to (1:1:1) regime (branches $B_2$ and $B_3$) and (2:1) regime (branches $B_5$ and $B_6$), merge and disappear resulting from the saddle-node bifurcation at $K \approx 7.023$ and $K \approx 7.042$ respectively. {For given values of the parameters, the obtained numerical results are in good agreement with the analytical theory described in Sec.~\ref{sec:Stability_Out_Of_Phase}.}
\par
On Fig.~\ref{fig4} it is to see that with the increasing values of $\lambda$ the stability region length of (1:1:1) regime becomes longer and the length of stability region for (2:1) regime becomes shorter.
\subsection{Chaotic dynamics} \label{sec:Chaos}
\begin{figure}[h!]
	\centering
	\includegraphics[width=\columnwidth]{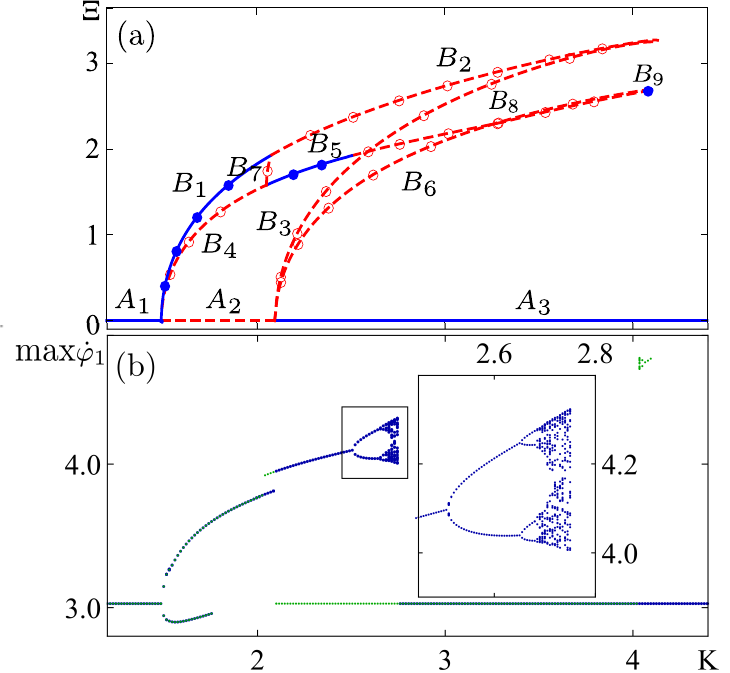}
	\caption{~(Color online) (a) Bifurcation diagram of synchronous rotational regimes of the system~\eqref{eq:EqPhi1} at $N=3$. (b) Local maxima of $\dot{\varphi}_1$. Blue dots -- the continuation of dynamical regime from the left to the right. Green dots -- the continuation of dynamical regime from the right to the left. Parameters: $\gamma = 0.8$, $\lambda = 0.3$.}
	\label{fig_gamma=0_8_lambda=0_3}
\end{figure}

Next let us consider the case with a lower value of the external force $\gamma = 0.8$.
Then at lower values of $\lambda$ the bifurcations of $2\pi$- and $4 \pi$-periodic regimes qualitatively coincide with the above described $\gamma = 0.97$.
With the further increasing damping parameter from branch $B_5$ of the solitary regimes motions with bigger periods appear resulting from doubling period bifurcations, and it leads to the appearance of chaotic dynamics.
Let us describe the same scenario for the case $\lambda = 0.3$ (Fig.~\ref{fig_gamma=0_8_lambda=0_3}).
Here at the increasing $K$ the (2:1) solitary regime (branch $B_5$) loses stability (branch $B_8$) at $K \approx 2.507$ resulting from period doubling bifurcation.
Further after cascade of period doubling bifurcations the (2:1) solitary regime becomes chaotic.
Branch $B_9$ corresponds to the stable (2:1) regime.
Similarly at a lower $K$ parameter at $K \approx 4.063$ the regime loses stability due to period-doubling bifurcation, what leads to the appearance of chaotic attractor. That exists at $2.7 \lesssim K < 2.754$. At $K > 2.754 $ the in-phase rotation becomes stable again.

\par
\begin{figure}[h!]
	\centering
	\includegraphics[width=\columnwidth]{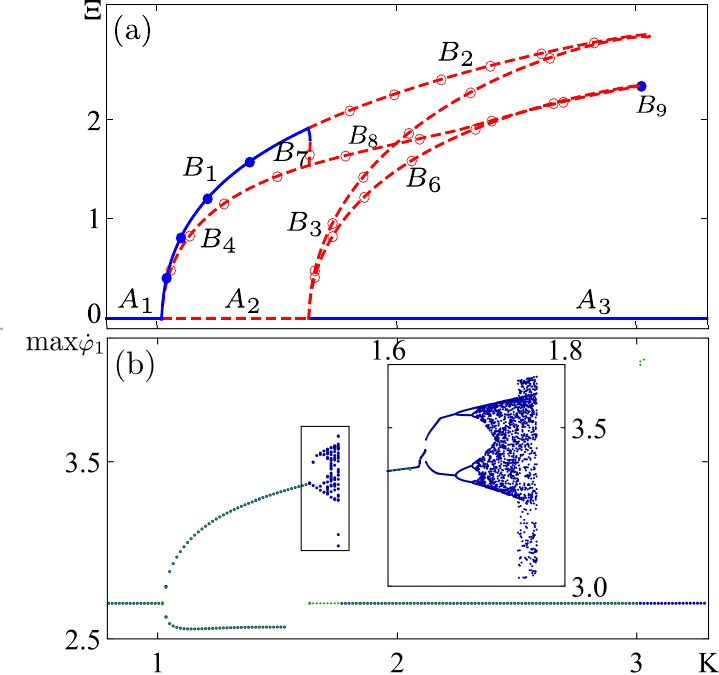}
	\caption{(Color online) Same as Fig.~\ref{fig_gamma=0_8_lambda=0_3}, but for $\gamma = 0.8$, $\lambda = 0.35$.}
	\label{fig_gamma=0_8_lambda=0_35}
\end{figure}
\begin{figure}[h!]
	\centering
	\includegraphics[width=\columnwidth]{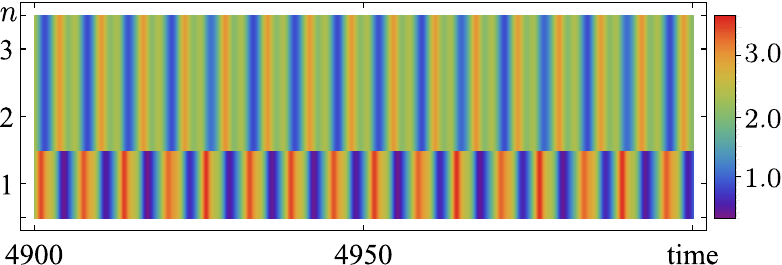}
	\caption{(Color online) Time dynamics of instantaneous frequencies $\dot{\varphi}_{n}$ ($n=1,2,3$) of the three pendulums in the system~\eqref{eq:EqPhi1} for $N=3$. Chaotic solitary (2:1) regime. Parameters: $\gamma = 0.8$, $\lambda = 0.35$, $K = 1.76$.}
	\label{fig_chaotic_chimera}
\end{figure}

Next at the increasing $\lambda$ the length of the stable branch of the (2:1) solitary state $B_5$ decreases, and it disappears.
Fig.~\ref{fig_gamma=0_8_lambda=0_35} shows the bifurcation diagrams of synchronous regimes for $\lambda = 0.35$. In this case chaotic solitary regime (Fig.~\ref{fig_chaotic_chimera}) is realized resulting from the cascade of period doubling bifurcations at $1.7 < K < 1.77$ and $3.0< K < 3.01$.
However, at $1.77 < K < 3.01$ the chaotic motion is unstable because of the chaotic attractor crisis, and in-phase rotation is to see.
\par
\begin{figure}[h!]
	\includegraphics[width=\columnwidth]{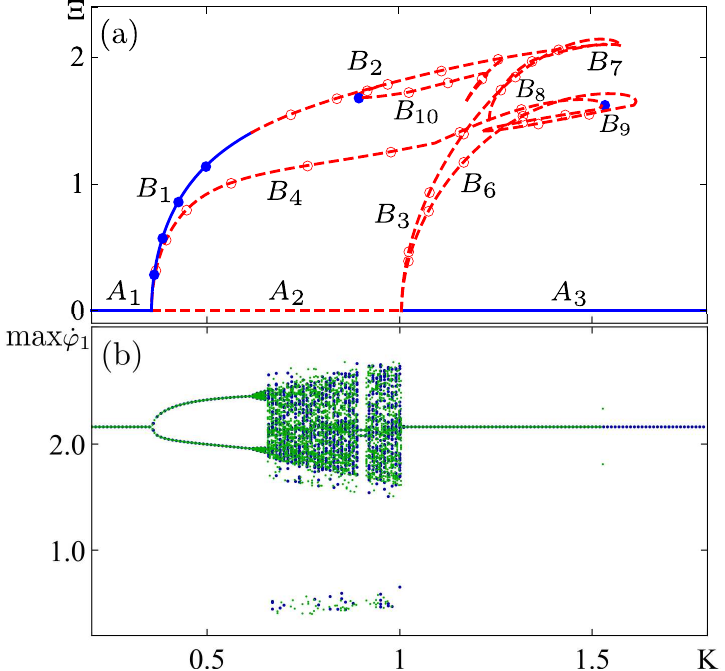}
	\caption{(Color online) Same as Fig.~\ref{fig_gamma=0_8_lambda=0_3}, but for $\gamma = 0.8$, $\lambda = 0.5$.}
	\label{fig_gamma=0_8_lambda=0_50}
\end{figure}
\begin{figure}[h!]
	\centering
	\includegraphics[width=\columnwidth]{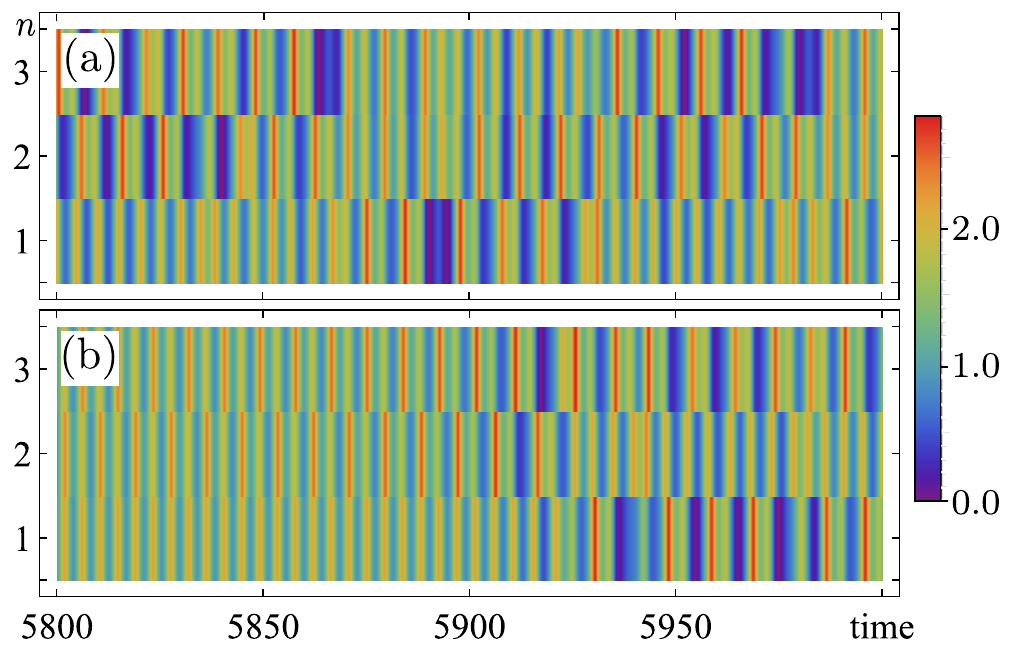}
	\caption{(Color online) Time dynamics of instantaneous frequencies $\dot{\varphi}_{n}$ ($n=1,2,3$) of the three pendulums in the system~\eqref{eq:EqPhi1} for $N=3$. (a) Chaotic (1:1:1) regime at $K = 0.95$. (b) Chaotic (1:1:1) regime with (1:1:1) and (3:0) intermittency at $K = 1.0$. Parameters: $\gamma = 0.8$, $\lambda = 0.5$.}
	\label{fig_chaotic_nochimera}
\end{figure}

The next case is for $\lambda = 0.5$.
At the increasing $K$ (1:1:1) motion (branch $B_1$) loses stability ($K \approx 0.618$) resulting from the Neimark-Sacker bifurcation, herewith quasiperiodic (1:1:1) motion appears.
At $K \approx 0.656$ this quasiperiodic regime becomes chaotic after torus destruction bifurcation (Fig.~\ref{fig_chaotic_nochimera}a).
At $0.89 < K < 1.272$ there is branch $B_{10}$ of $4\pi$-periodic motions, which merges with branch $B_2$ at $K \approx 0.956$ and $K \approx 1.272$ resulting from pitchfork bifurcations.
At $0.89 < K <0.907$ this branch has a stable region which corresponds to "stability window" when the chaos in system~\eqref{eq:EqPhi1} is not observed (Fig.~\ref{fig_gamma=0_8_lambda=0_50}b).
Further at the increasing $K$ the regime is realized, when time intervals, at which pendulums' phases $\varphi_1 \approx \varphi_2 \approx \varphi_3$ alternate with intervals, where $\varphi_1$, $\varphi_2$ and $\varphi_3$ do not coincide (Fig.~\ref{fig_chaotic_nochimera}b), i.e. there is an intermittency of chaotic oscillations (3:0) and (1:1:1).
If $K \approx 1.007$, the in-phase regime becomes stable and chaotic oscillations are not realized, and herewith the system shows the in-phase rotations.

\section{Rotational states in larger ensembles}\label{sec:N_4}
\begin{figure}[h!]
	\centering
	\includegraphics[width=\columnwidth]{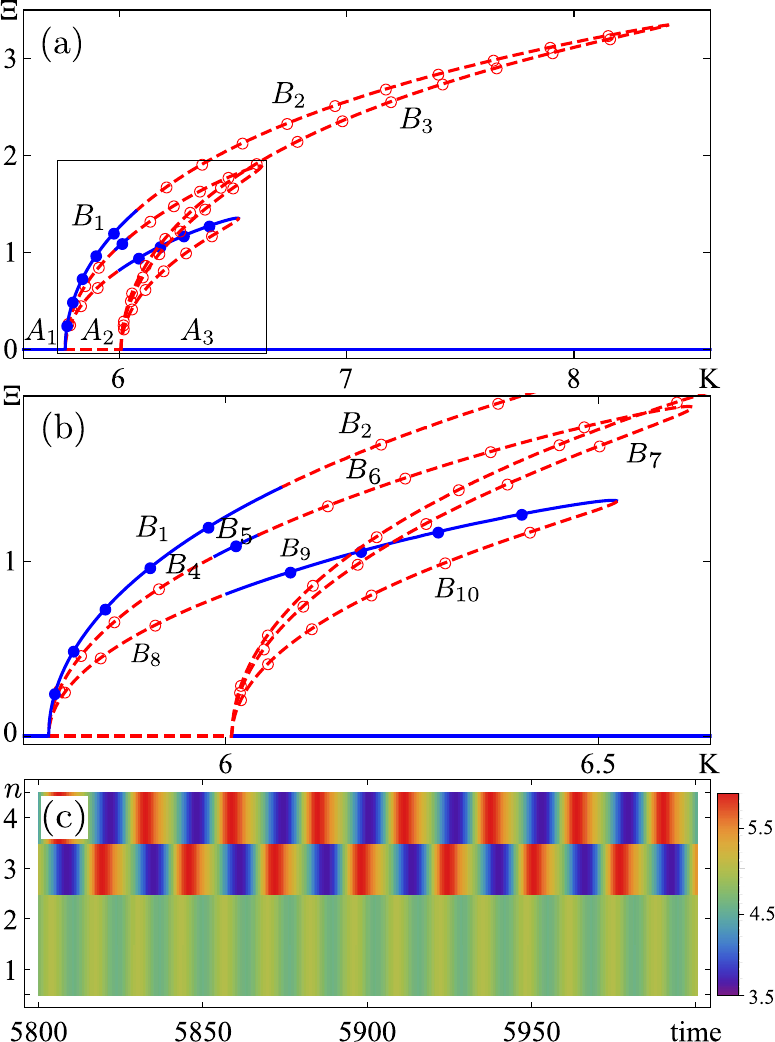}
	\caption{(Color online) (a), (b) Bifurcation diagram of synchronous rotational regimes of the system~\eqref{eq:EqPhi1} for $N=4$. (b) Time dynamics of instantaneous frequencies $\dot{\varphi}_{n}$ ($n=1,2,3,4$) of the four pendulums in the system~\eqref{eq:EqPhi1} for $N=4$. (c) Regular cluster (2:1:1) regime at $K = 0.6$. Parameters: $\gamma = 0.97$, $\lambda = 0.2$.}
	\label{fig_N_4}
\end{figure}

In this chapter we present a variant of the development of the in-phase rotational motion's instability for the system~\eqref{eq:EqPhi1} at $N>3$.
Let us consider the case $N=4$. The boundaries of the stability region of the in-phase rotation are defined by expressions~\eqref{eq_alpha0_K12_1} according to asymptotic theory developed in Section~\ref{sec:MainSection} for $\lambda \ll \gamma$. Next applying our analytical approach it can be shown there are three out-of-phase rotation motions: $(2:2)$ and $(3:1)$ two-clusters regimes and also $(2:1:1)$ cluster state. What is more according to developed theory this cluster state is stable for $K_{c1}^{(2:1:1)}<K<K_{c2}^{(2:1:1)}$, where
\begin{equation}
\begin{gathered}
K_{c1}^{(2:1:1)} = \dfrac{\gamma^2}{4 \lambda^2} + \dfrac{\sqrt{1-\gamma^2}}{4} + O(\lambda),\\
K_{c2}^{(2:1:1)} = \dfrac{\gamma^2}{4 \lambda^2} + \dfrac{\sqrt{1-\gamma^2}}{2} + O(\lambda).
\label{eq_stability_2_1_1}
\end{gathered}
\end{equation}
For example let us numerically and analytically describe system~\eqref{eq:EqPhi1} with parameters $\gamma=0.97$, $\lambda=0.2$ (Fig.~\ref{fig_N_4}). Direct numerical simulations show that for this values of coupling parameter $K$ in-phase periodic motion undergoes period doubling bifurcation. While out-of-phase (2:2), (2:1:1) and (3:1) regimes are generated (branches $B_1$, $B_2$ and $B_3$, respectively). Fig.~\ref{fig_N_4}(b) shows (2:1:1) regime, when two pendulums form an in-phase cluster and the others demonstrate out-of-phase dynamics. We note that in this case the range of parameter $K$ exists where out-of-phase (2:2), (2:1:1) and (3:1) regimes coexist and are stable, i.e. multistability is observed. In particular it is found from numerical calculations that for values $K_{c1}^{(2:1:1)} < K < K_{c2}^{(2:1:1)}$, where $K_{c1}^{(2:1:1)} \approx 5.9859$ and $K_{c2}^{(2:1:1)} \approx 6.0443$, (2:1:1) regime is stable. Corresponding theoretical values for the boundaries $K_{c1}^{(2:1:1)}$, $K_{c2}^{(2:1:1)}$ of the stability region following from the Eqs.~\eqref{eq_stability_2_1_1} are  $K_{c1}^{(2:1:1)} \approx 5.9414$ and $K_{c2}^{(2:1:1)} \approx 6.0022$.
\par
{We have analyzed the dynamics of ensembles with other number of elements. It was found that in the process of development of the in-phase ($N:0$) mode instability with an increase in the coupling strength $K$, a stable two-cluster state $(N_1:N_2)$ is first observed. Noteworthy, in such case the number of elements $N_1$ and $N_2$ in each cluster is approximately equal to each other. This state becomes unstable with further increase of the parameter $K$, and other cluster modes are realized. In all cases the sequence of cluster mode type is finished by solitary (two-cluster) state $(N-1:1)$.}
\par
\begin{figure}[h!]
	\centering
	\includegraphics[width=\columnwidth]{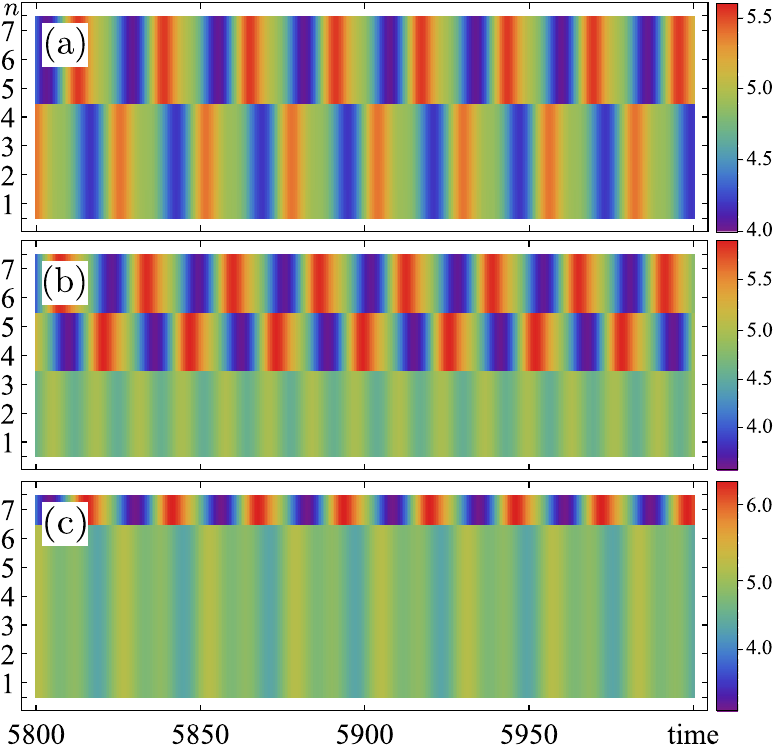}
	\caption{(Color online) Time dynamics of instantaneous frequencies $\dot{\varphi}_{n}$ ($n=1,2,\dots, 7$) of the pendulums in the system~\eqref{eq:EqPhi1} for $N=7$. (a) Cluster (4:3) regime at $K = 5.85$. (b) Cluster (3:2:2) regime at $K = 6.0$. (c) Solitary state (6:1) at $K = 6.1$. Parameters: $\gamma = 0.97$, $\lambda = 0.2$.}
	\vspace{-.1in}
	\label{fig_N_7}
\end{figure}

{We demonstrate this observation for the case $N=7$ pendulums with parameters $\gamma=0.97$ and $\lambda=0.2$. Due to in-phase rotation mode instability development the stable (4:3) cluster regime arises (Fig.~\ref{fig_N_7}a). Further there is a sequence of stable cluster modes of other types, for example (3:2:2) (Fig.~\ref{fig_N_7}b). Solitary state (6:1) is observed last in a sequence of stable cluster modes (Fig.~\ref{fig_N_7}c).}
\vspace{-.1in}
%----------%----------%----------%----------%----------%
\section{Conclusion}\label{sec:Conclusion}
We have studied the dynamics of an ensemble of globally coupled identical pendulums. A relatively simple model demonstrates a big variety of regular and chaotic in-phase and out-of-phase cluster regimes. We found and theoretically approved self-induced parametric instability of symmetric in-phase rotation regime.
It is shown, that in the system with the growth of coupling strength the generation of out-of-phase rotation periodic motions occurs resulting from period doubling bifurcation.
\par
Note that there is one instability region, where different out-of-phase regimes can be realized. {It describes an approach that allows to build asymptotic expansion for rotational modes and to investigate their stability in low dissipation limit. The developed method was applied in the case of a system consisting of three elements. The results of direct numerical modeling are well correlated with theoretical estimates of the boundaries of the existence and stability of rotational regimes, including solitary states, in the case of small dissipation.} Bistability of in-phase and out-of-phase rotational periodic regimes can also be observed.
\par
{It is also demonstrated that at larger values of dissipation parameter chaotic regimes and intermittency can be realized. The several examples of cluster modes, as well as analysis of their stability, were demonstrated in the case of $N=4$ and $N=7$.}
%------------------------------------------------------------------------------%
\acknowledgments
	Authors acknowledge A.O.~Kazakov, G.V.~Osipov and A.S.~Pikovsky for valuable advices and fruitful discussion.
	Results presented in Section~\ref{sec:MainSection} were supported by the RFBR grant No.~17-32-50096.
	Results presented in Section~\ref{sec:LimitCycles} were supported by the RSF grant No.~19-12-00367. Results presented in \ref{sec:N_4} were supported by the RFBR grant No.~19-52-12053.
%------------------------------------------------------------------------------%
\appendix


\section{Regimes (2:1) and (1:1:1) stability analysis}\label{sec:Stability_Analysis}
%------------------------------------------------------------------------------%
\subsection{Regime 2:1}\label{sec:Regime_2_1}
%------------------------------------------------------------------------------%
{Let us consider solitary state $(2:1)$, when two clusters exist: $\varphi_1(t) = \varphi_2(t) = \phi_1(t)$, $ \varphi_3(t) = \phi_2(t)$. We construct an asymptotic expansion for $\phi_1(t)$ and $\phi_1(t)$ in the case of $\lambda\!\ll\!\gamma$.} Applying coupling strength $K = \dfrac{\gamma^2}{4 \lambda^2} + \Delta K$, where the first summand $\dfrac{\gamma^2}{4 \lambda^2}$, to the first two leading orders, determines the middle of the coupling strength $K$ values range, for which the in-phase periodic motion becomes unstable (see Eq.~\ref{eq_alpha0_K12_1}), $\Delta K$ -- deviation from this value. Considering that $\phi_1^{(0)} = \phi_2^{(0)}$ and assuming for simplicity initial condition $\left(\phi_{1} - \phi_{2}\right)' |_{t=0} = 0$, we construct two asymptotic solutions $\phi_m^{+}$ and $\phi_m^{-}$ using methods described above for the case $\lambda \ll \gamma$:
\begin{equation}
\begin{aligned}
\phi_{1}^{\pm}(\tau) & = 2 \tau - \arccos(\pm \gamma_1) + \frac{2}{3}\frac{\lambda}{\gamma} A_2^{\pm}\cos \tau \\&
+ \frac{\lambda^2}{\gamma^2} \left[ \frac{\gamma}{24 \gamma_1} \left(36 \gamma_1 \mp \left(A_2^{\pm}\right)^2\right) - \gamma \cos 2 \tau \pm \gamma_1 \sin 2 \tau \right] \\ &+ A_2^{\pm}\frac{\lambda^3}{\gamma^3} \Biggl[ \pm \frac{16}{15} \gamma_1 + B_2^{\pm} \cos \tau - \frac{\gamma}{2} \sin \tau \\ &- \frac{1}{72} \left(\left(A_2^{\pm}\right)^2 \mp 12 \gamma_1\right) \cos 3 \tau  + \frac{\gamma}{6} \sin 3 \tau \Biggr] +  O(\lambda^4),\\
\phi_2^{\pm}(\tau) & = 2 \tau - \arccos(\pm \gamma_1) - \frac{4}{3}\frac{\lambda}{\gamma} A_2^{\pm}\cos \tau \\ & + \frac{\lambda^2}{\gamma^2} \left[ \frac{\gamma}{24 \gamma_1} \left(36 \gamma_1 \mp \left(A_2^{\pm}\right)^2\right) - \gamma \cos 2 \tau \pm \gamma_1 \sin 2 \tau \right] \\
& + A_2^{\pm} \frac{\lambda^3}{\gamma^3} \Biggl[\pm \frac{16}{15} \gamma_1 - 2 B_2^{\pm} \cos \tau + \gamma \sin \tau \\ & + \frac{1}{36} \left(\left(A_2^{\pm}\right)^2 \mp 12 \gamma_1\right) \cos 3 \tau - \frac{\gamma}{3} \sin 3 \tau \Biggr] +  O(\lambda^4),
\end{aligned}
\label{eq:2_1_phi_1_2_3}
\end{equation}
where
\begin{equation}
\begin{gathered}
A_2^{\pm} = 6 \sqrt{\frac{2}{5} \left(\Delta K \pm \frac{\gamma_1}{2}\right) },
\end{gathered}
\label{eq:A2pm}
\end{equation}
\begin{equation}
\begin{gathered}
B_2^{\pm} = -\frac{3}{5 \left(A_2^{\pm}\right)^2} - \frac{71}{720} \left(A_2^{\pm}\right)^2 \pm \left(\frac{7}{15} \gamma_1 - \frac{\gamma^2}{10 \gamma_1}\right),
\end{gathered}
\label{eq:B2pm}
\end{equation}
\begin{multline}
\tau = \bigg\{ \frac{\gamma}{2 \lambda} - \frac{\lambda}{\gamma} \frac{\left(A_2^{\pm}\right)^2}{18} \\
- \frac{\lambda^3}{\gamma^3} \left[ \frac{1}{4} + \frac{\left(A_2^{\pm}\right)^2}{18} \left(3 B_2^{\pm} - \frac{\left(A_2^{\pm}\right)^2}{9}\right) \right] + O(\lambda^5) \bigg\} t,
\label{eq:2_1_tau}
\end{multline}	
\begin{equation}
\gamma_1 = \sqrt{1 - \gamma^2}.
\end{equation}
{Thus there are two $(2:1)$ solitary states.}
Solution $\phi_m^{+}$ corresponds to the $4 \pi$-periodic regime, arising as a result of direct period doubling bifurcation from in-phase rotation $\phi(t)$ for $K=K_1$. Similarly, solution $\phi_m^{-}$ corresponds to $4 \pi$-periodic regime, arising as a result of inverse period doubling bifurcation for $K=K_2$.

Further, using the method described in Section \ref{sec:ParametricInstability}, we study the stability of the obtained solutions $\phi_m^{\pm}$. For the internal stability (see Eq.~\eqref{eq:EqTildeEta}) of the first solution $\phi_m^{+}$ we obtain
\begin{equation}
\begin{aligned}
\Lambda_1^{in} &= 0,\\
\Lambda_2^{in} &= - \lambda,\\
\Lambda_{3,4}^{in} &= - \frac{\lambda}{2} \left(1 \pm \sqrt{1 - \frac{\gamma_1}{\gamma^2} \left(A_2^{+}\right)^2}\right) + O(\lambda^2).
\end{aligned}
\label{eq:2_1_Lambda_in_p}
\end{equation}
Similarly, for the second one $\phi_m^{-}$:
\begin{equation}
\begin{aligned}
\Lambda_1^{in} &= 0,\\
\Lambda_2^{in} &= - \lambda,\\
\Lambda_{3,4}^{in} &= - \frac{\lambda}{2} \left(1 \pm \sqrt{1 + \frac{\gamma_1}{\gamma^2} \left(A_2^{-}\right)^2}\right) + O(\lambda^2).
\end{aligned}
\label{eq:2_1_Lambda_in_m}
\end{equation}

For the two-pendulums cluster external stability (see Eq.~\eqref{eq:EqTildeXi}) of the motion $\phi_m^{+}$ we have the following characteristic exponents:
\begin{equation}
\begin{aligned}
\Lambda_1^{ex} & = \lambda^3 \frac{\left(A_2^{+}\right)^4}{72 \gamma^4} \left(\gamma_1 - \frac{\left(A_2^{+}\right)^2}{12}\right) + O(\lambda^4),\\
\Lambda_2^{ex} & = -\lambda - \lambda^3 \frac{\left(A_2^{+}\right)^4}{72 \gamma^4} \left(\gamma_1 - \frac{\left(A_2^{+}\right)^2}{12}\right) + O(\lambda^4).
\end{aligned}
\label{eq:2_1_Lambda_ex_p}
\end{equation}
And for $\phi_m^{-}$:
\begin{equation}
\begin{aligned}
\Lambda_1^{ex} &= - \lambda^3 \frac{\left(A_2^{-}\right)^4}{72 \gamma^4} \left(\gamma_1 + \frac{\left(A_2^{-}\right)^2}{12}\right) + O(\lambda^4),\\
\Lambda_2^{ex} &= - \lambda + \lambda^3 \frac{\left(A_2^{-}\right)^4}{72 \gamma^4} \left(\gamma_1 + \frac{\left(A_2^{-}\right)^2}{12}\right) + O(\lambda^4).
\end{aligned}
\label{eq:2_1_Lambda_ex_m}
\end{equation}

%------------------------------------------------------------------------------%
\subsection{Regime 1:1:1}\label{sec:Regime_1_1_1}
%------------------------------------------------------------------------------%
The methods used to investigate stability of the regime (1:1:1) are completely analogous. Therefore, this section contains only brief description of the main results about the stability of the (1:1:1) motion.

Expressing $K$ in term of $\Delta K$ again, considering that $\phi_1^{(0)} = \phi_2^{(0)} = \phi_3^{(0)}$ and assuming initial conditions $\left(\phi_{1} - \phi_{3}\right)' |_{t=0}=~0$, we get two solutions $\phi_m^{+}$ and $\phi_m^{-}$:
\begin{equation}
\begin{aligned}
\phi_1^{\pm}(\tau) &= 2 \tau - \arccos(\pm \gamma_1) + A_3^{\pm}\frac{\lambda}{\gamma}\cos \tau\\ & + \frac{\lambda^2}{\gamma^2} \left[ \frac{\gamma}{32 \gamma_1} \left(48 \gamma_1 \mp \left(A_3^{\pm}\right)^2\right) - \gamma \cos 2 \tau \pm \gamma_1 \sin 2 \tau \right]\\
& + A_3^{\pm}\frac{\lambda^3}{\gamma^3} \Biggl[B_3^{\pm} \cos \tau - \frac{3}{4} \gamma \sin \tau\\ & - \frac{1}{64} \left(\left(A_3^{\pm}\right)^2 \mp 16 \gamma_1\right) \cos 3 \tau + \frac{\gamma}{4} \sin 3 \tau \Biggr] +  O(\lambda^4),\\
\phi_2^{\pm}(\tau) & = 2 \tau - \arccos(\pm \gamma_1)  + \frac{\lambda^2}{\gamma^2} \Biggl[ \frac{\gamma}{32 \gamma_1} \left(48 \gamma_1 \mp \left(A_3^{\pm}\right)^2\right)\\ & - \gamma \cos 2 \tau \pm \gamma_1 \sin 2 \tau \Biggr] +  O(\lambda^4),\\			
\phi_3^{\pm}(\tau) & = 2 \tau - \arccos(\pm \gamma_1) - A_3^{\pm}\frac{\lambda}{\gamma}\cos \tau\\ & + \frac{\lambda^2}{\gamma^2} \Biggl[ \frac{\gamma}{32 \gamma_1} \left(48 \gamma_1 \mp \left(A_3^{\pm}\right)^2\right) - \gamma \cos 2 \tau \pm \gamma_1 \sin 2 \tau \Biggr] \\ 
& - A_3^{\pm}\frac{\lambda^3}{\gamma^3} \Biggl[B_3^{\pm} \cos \tau - \frac{3}{4} \gamma \sin \tau\\ & - \frac{1}{64} \left(\left(A_3^{\pm}\right)^2 \mp 16 \gamma_1\right) \cos 3 \tau + \frac{\gamma}{4} \sin 3 \tau \Biggr] +  O(\lambda^4),
\end{aligned}
\label{eq:1_1_1_phi_1_2_3}
\end{equation}
where
\begin{equation}
A_3^{\pm} = 4 \sqrt{\frac{6}{5} \left(\Delta K \pm \frac{\gamma_1}{2}\right) },
\label{eq:A3pm}
\end{equation}
\begin{equation}
B_3^{\pm} = -\frac{6}{5 \left(A_3^{\pm}\right)^2} - \frac{11}{128} \left(A_3^{\pm}\right)^2 \pm \left(\frac{7}{10} \gamma_1 - \frac{3 \gamma^2}{20 \gamma_1}\right),
\label{eq:B3pm}
\end{equation}
\begin{multline}
\tau = \bigg\{ \frac{\gamma}{2 \lambda} - \frac{\lambda}{\gamma} \frac{\left(A_3^{\pm}\right)^2}{24} \\- \frac{\lambda^3}{\gamma^3} \left[ \frac{1}{4} + \frac{\left(A_3^{\pm}\right)^2}{12} \left(B_3^{\pm} - \frac{\left(A_3^{\pm}\right)^2}{24}\right) \right] + O(\lambda^5) \bigg\} t.
\label{eq:1_1_1_tau}
\end{multline}

In the case of (1:1:1) regime there are only one-pendulum clusters. For this $\phi_m^{+}$ solution characteristic exponents are
\begin{equation}
\begin{aligned}
\Lambda_1^{in} &= 0,\\
\Lambda_2^{in} &= -\lambda,\\
\Lambda_3^{in} &= - \lambda^3 \frac{\left(A_3^{+}\right)^4}{128 \gamma^4} \left(\gamma_1 - \frac{\left(A_3^{+}\right)^2}{16}\right) + O(\lambda^4),\\
\Lambda_4^{in} &= -\lambda + \lambda^3 \frac{\left(A_3^{+}\right)^4}{128 \gamma^4} \left(\gamma_1 - \frac{\left(A_3^{+}\right)^2}{16}\right) + O(\lambda^4),\\
\Lambda_{5,6}^{in} &= -\frac{\lambda}{2} \left(1 \pm \sqrt{1 - \frac{3 \gamma_1}{4 \gamma^2} \left(A_3^{+}\right)^2}\right) + O(\lambda^2).
\end{aligned}
\label{eq:1_1_1_Lambda_ex_p}
\end{equation}

Respectively for $\phi_m^{-}$:
\begin{equation}
\begin{aligned}
\Lambda_1^{in} &= 0,\\
\Lambda_2^{in} &= -\lambda,\\
\Lambda_3^{in} &= \lambda^3 \frac{\left(A_3^{-}\right)^4}{128 \gamma^4} \left(\gamma_1 + \frac{\left(A_3^{-}\right)^2}{16}\right) + O(\lambda^4),\\
\Lambda_4^{in} &= -\lambda - \lambda^3 \frac{\left(A_3^{-}\right)^4}{128 \gamma^4} \left(\gamma_1 + \frac{\left(A_3^{-}\right)^2}{16}\right) + O(\lambda^4),\\
\Lambda_{5,6}^{in} &= -\frac{\lambda}{2} \left(1 \pm \sqrt{1 + \frac{3 \gamma_1}{4 \gamma^2} \left(A_3^{-}\right)^2}\right) + O(\lambda^2).
\end{aligned}
\label{eq:1_1_1_Lambda_ex_m}
\end{equation}

%------------------------------------------------------------------------------%
\section{Methods for calculation of periodic motions and their stability}\label{sec:Numerical_Setup}
%------------------------------------------------------------------------------%
The theoretical analysis above allows us to describe the initial stage of the discussed instability of the synchronous rotation mode. One also can find all intervals of values of the coupling coefficient $K$, for which the development of the set-induced parametric instability is possible, and estimate the boundaries of these ranges with rather good accuracy. The direct numerical simulations of an initial value problem for the dynamical system~(\ref{eq:EqPhi1}) give us general ideas about the evolution in time of the ensemble of coupled pendulums and the nonlinear stage of the developed instability. In order to connect and complete these two pictures, we also identify periodic rotations and explore their parametric continuation within the framework of the model~(\ref{eq:EqPhi1}). To this end, taking into account that the $\varphi_{n}\left(t\right)$ is determined in the range from $-\pi$ to $\pi$ and using the property of closure of the considered trajectories in the phase space $\left\{\varphi_{n}\left(t\right),\dot{\varphi}_{n}\left(t\right)\right\}$, we construct
the Poincar\'{e} map and employ the Newton-Raphson algorithm to find a fixed point there and a period $T$ of motion along a corresponding trajectory for each given set of parameters $\lambda$, $\gamma$ and $K$~\cite{BifurcationTheory}.

The main idea of this method is as follows. Each of the solutions $\phi_{n}\!\left(t\right)$ we are interested in is primarily characterized by its period $T$ (which is strictly speaking unknown and is to be defined at the end of numerical computations) and the number $m$ of changes of phases $\varphi_{n}\!\left(t\right)$ by $2\pi$ during the period $T$. Hence, the Poincar\'{e} map $\bigl\{\varphi_{n}\left(0\right),\dot{\varphi}_{n}\left(0\right)\bigr\}\to\bigl\{\varphi_{n}\left(T\right)-2\pi{m},\dot{\varphi}_{n}\left(T\right)\bigr\}$ has a fixed point corresponding to a trajectory $\bigl\{\phi_{n}\left(t\right),\dot{\phi}_{n}\left(t\right)\bigr\}$. Using this fact that $\phi_{n}\left(T\right)=\phi_{n}\left(0\right)+2\pi{m}$ and $\dot{\phi}_{n}\left(T\right)=\dot{\phi}_{n}\left(0\right)$, we construct the following system of equations
\vspace{-.1in}
%\begin{widetext}
\begin{multline}\label{eq:pmap}
\mathbf{P}\left(T,\bigl\{{\varphi_{0}}_{n},\dot{\varphi_{0}}_{n}\bigr\}\right)=
\begin{bmatrix}
\bigl\{\varphi_{n}\left(T,\bigl\{{\varphi_{0}}_{n},\dot{\varphi_{0}}_{n}\bigr\}\right)\bigr\}\\
\bigl\{\dot{\varphi}_{n}\left(T,\bigl\{{\varphi_{0}}_{n},\dot{\varphi_{0}}_{n}\bigr\}\right)\bigr\}
\end{bmatrix}\\-
\begin{bmatrix}
\bigl\{{\varphi_{0}}_{n}+2\pi{m}\bigr\}\\
\bigl\{\dot{\varphi_{0}}_{n}\bigr\}
\end{bmatrix}=0,
\vspace{-.1in}
\end{multline}
%\end{widetext}
where $\bigl\{\varphi_{n}\left(t\right),\dot{\varphi}_{n}\left(t\right)\bigr\}$ is the solution to Eqs.~(\ref{eq:EqPhi1}) with initial conditions $\bigl\{{\varphi_{0}}_{n},\dot{\varphi_{0}}_{n}\bigr\}$, i.e. $\bigl\{\varphi_{n}\left(0\right),\dot{\varphi}_{n}\left(0\right)\bigr\}=\bigl\{{\varphi_{0}}_{n},\dot{\varphi_{0}}_{n}\bigr\}$. Therefore, a periodic solution with period $T$ of Eqs.~(\ref{eq:EqPhi1}) will be a root to~(\ref{eq:pmap}). Because of the translational invariance symmetry (in time), we note that one value from the set $\bigl\{{\varphi_{0}}_{n}\bigr\}$ can always be taken to be zero without loss of generality. We use the Newton-Raphson algorithm~\cite{NumericalRecipes} to approximate the roots of $\mathbf{P}\left(T,\bigl\{{\varphi_{0}}_{n},\dot{\varphi_{0}}_{n}\bigr\}\right)$. It is also noteworthy that the Jacobian is $\widehat{\mathbf{J}}=\widehat{\mathbf{I}}-\widehat{\mathbf{Q}}\left(T\right)$, where $\widehat{\mathbf{I}}$ is the identical matrix and $\widehat{\mathbf{Q}}\left(T\right)$ is matrix obtained from the monodromy matrix $\widehat{\mathbf{M}}\left(T\right)$ (see its definition below) by replacing one of the columns by the vector of values of the right-hand sides of Eqs.~(\ref{eq:EqPhi1}) at the time $t=T$. As a result, we numerically obtain, with high precision, stable (above dynamically generated) and unstable rotational modes as exact time-periodic solutions of Eqs.~(\ref{eq:EqPhi1}). Continuing these solutions in value of the coupling strength $K$ within the interval of instability of in-phase rotational mode allows us to trace the entire family of nontrivial periodic motions and to analyze their bifurcations (see, e.g., Fig.~\ref{fig4}) and study in detail their bifurcations and a process of transition to chaos. This is one of the main goals of the paper.

The linear (spectral) stability of the arbitrary ($2\pi$-, $4\pi$-, $8\pi$- and etc.) periodic motions (on the cylinder) of the dynamical system~(\ref{eq:EqPhi1}) is investigated by means of a Floquet analysis, chiefly relying on numerical calculations (see, e.g.,~\cite{BifurcationTheory}). To this end, we add a small perturbation $\delta\varphi_{n}\!\left(t\right)$ to a given periodic solution $\phi_{n}\!\left(t\right)$ of Eqs.~(\ref{eq:EqPhi1}). The linearized equations satisfied to the first order in $\delta\varphi_{n}\!\left(t\right)$ read:
\vspace{-.1in}
\begin{equation}
\begin{gathered}
\delta\ddot{\varphi}_n + \lambda \delta\dot{\varphi}_n + \cos \phi_n(t) {\delta\varphi}_n \\
=\dfrac{K}{N}\sum_{\tilde{n}=1}^{N}\cos\left(\phi_{\tilde{n}}-\phi_n\right)\left({\delta\varphi}_{\tilde{n}} - {\delta\varphi}_n\right).
\end{gathered}
\vspace{-.1in}
\label{eq:FloquetAnalysis}
\end{equation}
The Floquet analysis of Eqs.~(\ref{eq:FloquetAnalysis}) can be performed due to the property of periodicity of the trajectory $\bigl\{\phi_{n}\left(t\right),\dot{\phi}_{n}\left(t\right)\bigr\}$. Therefore, the stability of the considered motion is defined by the spectrum of the Floquet operator (monodromy matrix) $\widehat{\mathbf{M}}\left(T\right)$ given by
\vspace{-.1in}
\begin{equation}
\begin{bmatrix}
\left\{\delta\varphi_{n}\left(T\right)\right\}\\
\left\{\delta\dot{\varphi}_{n}\left(T\right)\right\}
\end{bmatrix}=
\widehat{\mathbf{M}}
\begin{bmatrix}
\left\{\delta\varphi_{n}\left(0\right)\right\}\\
\left\{\delta\dot{\varphi}_{n}\left(0\right)\right\}
\end{bmatrix}.
\vspace{-.1in}
\end{equation}
The eigenvalues $\mu_{n'}=\exp\left(iq_{n'}\right)$ (here and below $n'=1,\ldots,2N$) of the matrix $\widehat{\mathbf{M}}\left(T\right)$ are dubbed the Floquet multipliers with being Floquet exponents $q_{n'}$ of the periodic solution $\phi_{n}\!\left(t\right)$. Because of the damping and the external force, only one property (implied by the real character of the monodromy) can be extracted for $\mu_{n'}$, which is that they are or appear in complex conjugated pairs. To examine the stability of each of the rotation motion under discussion, we compute their Floquet multipliers. If $\left|\mu_{n'}\right|\leq{1}$ for all $n'$, then the rotation mode is linearly stable. It is worth mentioning that one of the eigenvalues $\mu_{n'}$ must be strictly equal to one, because we investigate the stability of a periodic motion. Hence, using this fact it is possible to check that the trajectory $\bigl\{\phi_{n}\left(t\right),\dot{\phi}_{n}\left(t\right)\bigr\}$ we find numerically belongs to the family of periodic rotations (on the cylinder).
If at least one of Floquet multipliers $\mu_{n'}$ locates outside the unit circle in the complex plane, then the rotation mode is linearly unstable. Noteworthy, our calculations show that the most of the discussed periodic solutions $\varphi_{n}\left(t\right)$ exhibit a relatively strong instability due to a real Floquet multiplier $\mu_{n'}$ with an absolute value greater than one, i.e. $\left|\mu_{n'}\right|>1$.

As a characteristic of the degree of synchronization, we consider the value $\Xi$, which is the frequency lag of pendulums:
\vspace{-.1in}
\begin{equation}
\Xi = \dfrac{1}{N (N - 1)}\sum_{n_1, n_2 = 1}^N\max_{0 < t < T}|{\dot{\varphi}}_{n_1}(t) - {\dot{\varphi}}_{n_2}(t)|,
\vspace{-.1in}
\label{eq_Xi}
\end{equation}
where $T$ is the period of rotational mode. It follows from the definition (\ref{eq_Xi}) that $\Xi$ takes non-negative values, and $\Xi~=~0$ only in the case of in-phase mode. In the case of an out-of-phase regime, when there exists such a pair of pendulums that $\dot {\varphi}_{n_1}~\ne~\dot{\varphi}_{n_2}$, where ${n_1}$ and ${n_2}$ are the numbers of pendulums, $\Xi~>0$.


\begin{thebibliography}{09}


\bibitem{Pikovsky} A.~Pikovsky, M.~Rosenblum, and J.~Kurths, {``Synchronization. A Universal Concept in Nonlinear Sciences''} (Cambridge University Press, 2001).

\bibitem{Osipov} G.~V.~Osipov, J.~Kurths, and Ch.~Zhou, {``Synchronization in Oscillatory Networks''} (Springer Verlag: Berlin, 2007).

\bibitem{Afraimovich} V.~S.~Afraimovich, V.~I.~Nekorkin, G.~V.~Osipov, and V.~D.~Shalfeev, {``Stability, Structures and Chaos in Nonlinear Synchronization Networks''} (World Scientific, Singapore, 1994).

\bibitem{Mosekilde} E.~Mosekilde, Y.~Maistrenko and D.~Postnov, {``Chaotic Synchronization: Applications to Living Systems''} (World Scientific, Singapore, 2002).

\bibitem{Anishchenko} V.~Anishchenko, A.~Neiman, T.~Vadivasova,
V.~Astakhov, and L.~Schimansky-Geier, {``Nonlinear Dynamics of Chaotic and Stochastic Systems. Second edition: improved and enlarged''} (Springer Verlag: Berlin, 2007).

\bibitem{Balanov}  Al.~Balanov, N.~Janson, D.~Postnov, and O.~Sosnovtseva, {``Synchronization.	From Simple to Complex''} (Springer Verlag: Berlin, 2009).

\bibitem{Kecik} K.~Kecik, and J. Warminski, ``Dynamics of an Autoparametric Pendulum-Like System with a Nonlinear Semiactive Suspension'', Mathematical Problems in Engineering, 451047, 15 (2011).

\bibitem{Barone} A.~Barone, G.~Paterno, {``Physics and Applications of the Josephson Effect''} (John Wiley and Sons Inc., 1982).

\bibitem{Yakushevich} L.~V.~Yakushevich, {``Nonlinear Physics of DNA''} (2nd ed., Weinheim, Wiley-VCH, 2004).

%\bibitem{BraunKivshar} O.~M.~Braun and Yu.~S.~Kivshar, {``The Frenkel-Kontorova Model: Concepts, Methods, and Applications''} (Berlin, Springer, 2004).

%\bibitem{Yakushevich2011} L.~V.~Yakushevich, S.~Gapa, J.~Awrejcewicz, International Journal of Bifurcation and Chaos \textbf{21}, 3063 (2011).

\bibitem{PikovskyRosenblum2015} A.~Pikovsky and M.~Rosenblum, ``Dynamics of globally coupled oscillators: progress and perspectives'', Chaos {25}, 097616 (2015).

\bibitem{Kurths2016} F.~A. Rodrigues, T.~K.~D.Peron, P. Ji, and J. Kurths, ``Kuramoto model in complex networks'',  Phys. Rep. {610}, 1 (2016).

\bibitem{Kaneko1990} K.~Kaneko, ``Clustering, coding, switching, hierarchical ordering, and control in a network of chaotic elements'',  Physica D: Nonlinear Phenomena {41}, 137 (1990).

\bibitem{Okuda1993} K.~Okuda, ``Variety and generality of clustering in globally coupled oscillators'',  Physica D: Nonlinear Phenomena {63}, 424 (1993).

\bibitem{Nakagawa1994} N.~Nakagawa and Y.~Kuramoto, ``From collective oscillations to collective chaos in a globally coupled oscillator system'',  Physica D: Nonlinear Phenomena {75}, 74 (1994).

\bibitem{Schmidt2014} L.~Schmidt and K.~Krischer, ``Two-cluster solutions in an ensemble of generic limit-cycle oscillators with periodic self-forcing via the mean-field'',  Phys. Rev. E {90}, 042911 (2014).

\bibitem{Vanag2000} V.K.~Vanag, L.~Yang, M.~Dolnik, A.M.~Zhabotinsky, and I.R.~Epstein, ``Oscillatory cluster patterns in a homogeneous chemical system with global feedback'', Nature {406}, 389 (2000).

\bibitem{Mikhailov2006} A.S.~Mikhailov and K.~Showalter, ``Control of waves, patterns and turbulence in chemical systems'', Phys. Rep. {425}, 79 (2006).

\bibitem{Lin2004} A.L.~Lin, A.~Hagberg, E.~Meron, and H.L.~Swinney, ``Resonance tongues and patterns in periodically forced reaction-diffusion systems'', Phys. Rev. E {69}, 066217 (2004).

\bibitem{Kemeth2019} F.P.~Kemeth, S.W.~Haugland, and K.~Krischer, ``Cluster singularity: The unfolding of clustering behavior in globally coupled Stuart-Landau oscillators'', Chaos {29}, 023107 (2019).

\bibitem{Maistrenko2014} Y.~Maistrenko, B.~Penkovsky, and M.~Rosenblum, ``Solitary state at the edge of synchrony in ensembles with attractive and repulsive interactions'', Phys. Rev. E {89}, 060901(R) (2014).

\bibitem{Majhi2019} S.~Majhi, T.~Kapitaniak, and D.~Ghosh, ``Solitary states in multiplex networks owing to competing interactions'', Chaos {29}, 013108 (2019).

\bibitem{Mikhaylenko2019} M.~Mikhaylenko, L.~Ramlow, S.~Jalan, and A.~Zakharova, ``Weak multiplexing in neural networks: Switching between chimera and solitary states'', Chaos {29}, 023122 (2019).

\bibitem{Jaros2015} P.~Jaros, Y.~Maistrenko, and T.~Kapitaniak, ``Chimera states on the route from coherence to rotating waves'', Phys. Rev. E {91}, 022907 (2015).

\bibitem{Ji} P.~Ji, T.~Peron, F.~Rodrigues, and J.~Kurths, ``Low-dimensional behavior of Kuramoto model with inertia in complex networks'', Sci. Rep. {4}, 4783 (2014).

\bibitem{Ha} S.~Ha, Y.~Kim, and Z.~Li, ``Large-Time Dynamics of Kuramoto Oscillators under the Effects of Inertia and Frustration'', SIAM Journal on Applied Dynamical Systems, 13, 1, 466-492 (2014).

\bibitem{Belykh2016} I.~V.~Belykh, B.~N.~Brister, and V.~N.~Belykh, ``Bistability of patterns of synchrony in Kuramoto oscillators with inertia'', Chaos 26, 094822 (2016).

\bibitem{Komin} N.~Komin, and R.~Toral, ``Order parameter expansion and finite-size scaling study of coherent dynamics induced by quenched noise in the active rotator model'', Phys. Rev. E 82, 051127 (2010).

\bibitem{Daido} H.~Daido, ``Susceptibility of large populations of coupled oscillators'', Phys. Rev. E 91, 012925 (2015).

\bibitem{Lafuerza} L.~Lafuerza, P.~Colet, and R.~Toral, ``Nonuniversal Results Induced by Diversity Distribution in Coupled Excitable Systems'', Phys. Rev. Lett. 105, 084101 (2010).

\bibitem{Andronov} A.~A.~Andronov, A.~A.~Vitt, and S.~E.~Khaikin, in {``Adiwes International Series in Physics, Theory of Oscillators''} (Pergamon, 1966).

\bibitem{Tricomi1933} F. Tricomi, ``Integrazione di un' equazione differenziale presentatasi in elettrotecnica'', Ann. Scuolu Norm. Sup. Pisa {2}, l-20 (1933).

\bibitem{Belykh1977} V.~N.~Belykh, N.~F.~Pedersen, and O.~H.~Soerensen, ``Shunted-Josephson-junction model. I. The autonomous case'', Phys. Rev. B {16}, 4853 (1977).

\bibitem{Peng} Ji~Peng, T.~Peron, P.~Menck, F.~Rodrigues, and J.~Kurths. ``Cluster explosive synchronization in complex networks'', Phys. Rev. Lett.
110, 218701 (2013).

\bibitem{Strogatz} S.H.~Strogatz, ``Nonlinear Dynamics and Chaos. With Applications to Physics, Chemistry and Engineering'' (Reading, PA: Addison-Wesley, 1994).

\bibitem{Smirnov2016}  L.~A.~Smirnov, A.~K.~Kryukov, G.~V.~Osipov, and J.~Kurths, ``Bistability of rotational modes in a system of coupled pendulums'', Regul. Chaotic Dyn. {21}, 849-861 (2016).

\bibitem{Bolotov2019} M.~I.~Bolotov, V.~O.~Munyaev, A.~K.~Kryukov, L.~A.~Smirnov, and G.~V.~Osipov, ``Variety of rotation modes in a small chain of coupled pendulums'', Chaos 29, 033109 (2019).

\bibitem{Nayfeh} A.~H.~Nayfeh, {``Perturbation Methods''} (John Wiley, New York, 1973).

\bibitem{McLachlan} N.~W.~McLachlan, {``Theory and Application of Mathieu Functions''} (Clarendon Press, Oxford, 1947).

\bibitem{LyapunovExponents} A.~Pikovsky and A.~Politi, {``Lyapunov Exponents. A Tool to Explore Complex Dynamics''} (Cambridge University Press, 2016).

\bibitem{BifurcationTheory} Y.~A. Kuznetsov, {``Elements of Applied Bifurcation Theory''} (Springer, New York, 1995).

\bibitem{NumericalRecipes} W.~H. Press, S.~A. Teukolsky, W.~T. Vetterling, and B.~P. Flannery, {``Numerical Recipes: The Art of Scientific Computing''} (3rd ed., Cambridge University Press, New York, 2007).

\end{thebibliography}

	\appendix*

\end{document}